\documentclass{ws-mpla}
\usepackage[super]{cite}
\usepackage{graphicx}
\begin{document}

\markboth{M. Govender and S. Thirukkanesh}
{Causal heat flow in Bianchi type-V universe}

%%%%%%%%%%%%%%%%%%%%% Publisher's Area please ignore %%%%%%%%%%%%%%
\catchline{}{}{}{}{}
%%%%%%%%%%%%%%%%%%%%%%%%%%%%%%%%%%%%%%%%%%%%%%%%%%%%%%%%%%%%%%%%%%%

\title{CAUSAL HEAT FLOW IN BIANCHI TYPE-V UNIVERSE}

\author{\footnotesize M GOVENDER\footnote{
govenderm43@ukzn.ac.za}}

\address{Astrophysics and Cosmology Research Unit, School of Mathematics, Statistics and Computer Science, University of KwaZulu Natal, Private Bag X54001, Durban 4000, South Africa}

\author{\footnotesize S. THIRUKANNESH\footnote{thirukkanesh@yahoo.co.uk}}

\address{Department of Mathematics, Eastern University, Chenkalady, Sri Lanka.
}

\maketitle

\pub{Received (Day Month Year)}{Revised (Day Month Year)}

\begin{abstract}
In this paper we investigate the role of causal heat transport in
a spatially homogeneous, locally-rotationally symmetric Bianchi
type-V cosmological model. In particular, the causal temperature
profile of the cosmological fluid is obtained within the framework
of extended irreversible thermodynamics. We demonstrate that
relaxational effects can alter the temperature profile when the
cosmological fluid is out of hydrostatic equilibrium.
\end{abstract}

\ccode{PACS Nos.: include PACS Nos.}

\section{Introduction}	

While the standard Big Bang cosmological model has accounted for observations of homogeneity and isotropy
of the Universe on large scales there are still open questions regarding the identification of the dark energy
components making up the cosmic fluid \cite{roy1}. This has led to the pursuit
of more general models in which the geometry and the matter
content have drastically changed when compared to the standard FRW
cosmologies. To date the $\Lambda$CDM concordance model has proved
to be highly successful in accounting for all current observations
ranging from supernovae Ia, CMBR anisotropies,weak lensing, baryon
oscillations through to large-scale structure formation\cite{dunsby1}. There are various alternative cosmological models
ranging from inhomogeneous cosmologies with dissipative fluxes
\cite{inflation1,roy2}, singularity-free models
\cite{dadhich1,dadhich2}, emergent Universe models \cite{sailo1,pavon1} and the spatially homogeneous Bianchi models
\cite{prad,singh3}. It is claimed that these models
can account for many of the mechanisms leading to the current
state of the Universe such as inflation, particle production and
anisotropy in the infant Universe. The homogeneous and isotropic FRW cosmological models are particular cases of the Bianchi I, V and IX universes, depending on the constant curvature of the physical three-space, $t =$ constant. In particular, the Bianchi V universe is a simple generalisation of the negative curvature FRW models \cite{prad, singh3}. These alternative models can
represent particular epochs during the evolution of our Universe.
As pointed out by Ellis\cite{ellis1} the anisotropic Bianchi-type cosmologies
are worthy of attention even if current observations indicate that
our Universe is FLWR-like in nature. The observed isotropy of
the present Universe does not rule out the possibility of
dominant anisotropic effects in the early Universe.

The role of dissipation in inhomogeneous cosmological models have
been widely studied. Romano and Pavon studied the evolution of the
Bianchi type-I model with viscous dissipation within the framework
of extended irreversible thermodynamics\cite{romano}. They were
able to show that the Bianchi type-I cosmological model does not
asymptotically evolve into the Friedmann or de-Sitter phase. This
is mainly due to relaxational effects within the cosmological
fluid. Romano and Pavon utilised both the truncated and full causal thermodynamic theory to study the evolution of the Bianchi type-III cosmological models.
Their results show that there is rapid dissipation of the initial anisotropies leading to stable de Sitter solutions while the Friedmann ones are unstable \cite{romano2}. Causal heat transport in an inhomogeneous cosmological
model was investigated by Triginer and Pavon \cite{pavvo}. By
imposing a barotropic equation of state and by employing a heat
transport equation of Maxwell-Cattaneo form they were able obtain
more general behaviour of the scale factor and entropy production
for various spherically symmetric, inhomogeneous cosmological
models. Singh and Beesham \cite{bee1} investigated the effect of heat flow in a LRS Bianchi type-V universe with constant deceleration parameter.
They calculated the temperature distribution for the cosmological fluid by employing the Eckart transport equation for the heat flow.

In this paper we revisit the model investigated by Singh and Beesham with the view of highlighting the relaxational effects on the temperature distribution. To this end we employ a causal heat transport equation of Maxwell-Cattaneo form. By assuming that the relaxation time is inversely proportional to the inverse of the absolute value of the expansion of the cosmic fluid we are able to integrate the truncated heat transport equation to obtain the temperature profile. Our results show distinct differences between the causal and noncausal temperatures throughout the cosmic fluid.

\section{LRS Bianchi type-V cosmology}

The line element for locally-rotationally symmetric Bianchi type-V
cosmological model is given by\cite{bee1}
\begin{equation} \label{a1} ds^2 = -dt^2 + A^2dx^2 + e^{2x}B^2\left(dy^2 + dz^2\right),\,
\end{equation}
where $A = A(t)$ and $B = B(t)$ are metric functions yet to be
determined. The matter distribution for the cosmological fluid
interior is represented by the energy momentum tensor of an
imperfect fluid
\begin{equation}\label{mo}
T_{ab} = (\rho + p)u_a u_b + p g_{ab} + Q_a u_b
          + Q_b u_a,
\end{equation}
where $\rho$ is the energy density, $p$ is the pressure and $Q =
(Q^aQ_a)^{\frac{1}{2}}$ is the magnitude of the heat flux. The
fluid four--velocity ${\bf u}$ is comoving and is given by
\begin{equation}
u^a = \displaystyle \delta^{a}_0.\label{2'}
\end{equation}
The heat flow vector takes the form
\begin{equation}
Q^a = (0, Q^1, 0, 0), \label{2''}
\end{equation}
since $ Q^au_a = 0 $ and the heat is assumed to flow in the radial
direction. The fluid collapse rate $\Theta = u^a_{;a}$ of the
stellar model is given by
\begin{equation}
\Theta = \frac{\dot{A}}{A}+2\frac{\dot{B}}{B}. \label{2'''}
\end{equation}
The Einstein field equations reduce to
\begin{eqnarray}
\label{t4} \rho &=&
2 \frac{\dot{A}}{A}\frac{\dot{B}}{B} + \frac{\dot{B}^2}{B^2}- \frac{3}{A^2},\label{t4a} \\ \nonumber \\
p &=& \frac{1}{A^2}-\frac{\dot{B}^2}{B^2} -2 \frac{\ddot{B}}{B}, \label{t4b}  \\  \nonumber \\
p &=&\frac{1}{A^2}- \frac{\dot{A}}{A}\frac{\dot{B}}{B}-
\frac{\ddot{A}}{A} - \frac{\ddot{B}}{B},
\label{t4c}  \\ \nonumber \\
Q_1 &=& 2\left(\frac{\dot{B}}{B}- \frac{\dot{A}}{A}\right) ,
\label{t4d}
\end{eqnarray}
for the line element (\ref{a1}). The generalized mean Hubble
parameter $H$ is given by
\begin{equation}
\label{09}H=\frac{\dot{a}}{a}=\frac{1}{3}
\left(\frac{\dot{A}}{A}+2\frac{\dot{B}}{B}\right),
\end{equation}
where $a= (AB^2)^{1/3}$ is the average scale factor. The dot
denotes a derivative with respect to cosmic time $t$.

\section{Evolution of Hubble parameter}
Observations of the CMB and SNe Ia data point to an accelerating
universe ($q < 0$) where $q$ is the deceleration parameter.
Following Singh {et al} \cite{singh3} we assume that the Hubble
parameter is related to the average scale factor by
\begin{equation}
\label{10} H=la^{-n}=l(AB^2)^{-n/3},
\end{equation}
where $l(> 0)$ and $n(\geq 0)$ are constants, and
\begin{equation}
\label{11} n=q+1,
\end{equation}
where $H$ is defined as in eq. (\ref{09}) and $q$ the deceleration
parameter defined by
\begin{equation}
\label{12} q= -\frac{\ddot{a}a}{\dot{a}^2}.
\end{equation}
Using eqs. (\ref{10}) and (\ref{11}), the solution of eq.
(\ref{12}) gives the law of variation of average scale factor of
the form
\begin{equation}
\label{13}a=(nlt)^{1/n},
\end{equation}
for $n\neq 0$ and
\begin{equation}
\label{14} a=c \exp[lt],
\end{equation}
for $n=0$, where $c$ is the constant of integration. Here, in eq.
(\ref{13}), we have assumed that for $t=0$ the value $a=0$ so that
the constant of integration vanishes.

Now, from eqs. (\ref{t4b}) and (\ref{t4c}), we get
\begin{equation}
\label{15}\frac{\ddot{B}}{B}-\frac{\ddot{A}}{A}+\frac{\dot{B}^2}{B^2}
-\frac{\dot{A}}{A}\frac{\dot{B}}{B}=0.
\end{equation}
Integrating eq. (\ref{15}) and utilizing $a=(AB^2)^{1/3}$, the
metric functions $A$ and $B$ can be expressed as quadratures
\begin{eqnarray}
\label{16} A(t)&=& (d_1)^{-2/3} a \exp \left[- \frac{2k_1}{3}\int
a^{-3}dt\right],\\
\label{17} B(t)&=&(d_1)^{1/3} a \exp \left[ \frac{k_1}{3}\int
a^{-3}dt\right],
\end{eqnarray}
where $k_1$ and $d_1$ are the constants of integration.

The case $n \neq 0$ was considered by Singh and Beesham
\cite{bee1}. Our aim is to investigate relaxational effects when
the cosmic fluid leaves hydrostatic equilibrium. To this end we
consider the case $n = 0$ which is equivalent to $q = -1$ which
corresponds to inflation. Substituting eq. (\ref{14}) into eqs.
(\ref{16}) and (\ref{17}), the solution of the metric functions is
given by
\begin{eqnarray}
\label{23} A(t)&=& (d_1)^{-2/3} c \exp \left[lt+ \frac{2k_1}{3lc^3} \exp (-3lt)\right],\\
\label{24} B(t)&=&(d_1)^{1/3} c \exp \left[lt- \frac{k_1}{3lc^3}
\exp (-3lt) \right].
\end{eqnarray}
The  heat flow is given by
\begin{equation}
\label{25} Q_1 =\frac{2k_1}{c^3} \exp (-3lt).
\end{equation}
The energy density and pressure are respectively given by
\begin{eqnarray}
\label{26}\rho&=& 3l^2-\frac{1}{3} \frac{k_1^2}{c^6} \exp (-6lt) \nonumber\\
&& -3(d_1)^{4/3} c^{-2} \exp \left[
-2 \left(lt+ \frac{2k_1}{3lc^3} \exp [-3lt]\right)\right],\\
\label{27}p&=& -3l^2-\frac{1}{3} \frac{k_1^2}{c^6} \exp (-6lt) \nonumber\\
&& +(d_1)^{4/3} c^{-2} \exp \left[ -2 \left(lt+ \frac{2k_1}{3lc^3}
\exp [-3lt]\right)\right].
\end{eqnarray}

As pointed out by Singh and Beesham, the Universe as described by
this model starts evolving with constant kinematical and
thermodynamical parameters and maintains a constant expansion
rate. At late times this model mimicks an inflationary-like
behaviour with an equation of state $p = -\rho$. Inflation driven
by heat flux was demonstrated by Maartens {\em et
al}\cite{inflation1} in which they showed that the heat flux
serves to `balance' the decrease in energy density while the
pressure of the cosmic fluid steadily decreases. In order to
determine the deviation of the cosmic fluid from hydrostatic
equilibrium we calculate the covariant dimensionless ratio
\begin{equation}
\frac{|Q|}{\rho} =
2\frac{\sqrt{\frac{(d_1)^{4/3}k_1^2\exp[-\frac{\exp[-3ltk_1]-8lt}{3c^3l}]}{c^8}}}{-\frac{3(d_1)^{3/2}\exp[\frac{-\exp[-3ltk_1]-2lt}{3c^3l}]}{c^2}-
\frac{k_1^2\exp[-6lt]}{3c^6} + 3l^2},\end{equation} which for late
times decreases rapidly indicating that $|Q|$ decreases less
rapidly than $\rho$ during this epoch. Herrera {\em et al}
\cite{pavvo2} have shown that a certain parameter $\alpha$ defined
by
\[
\alpha = \frac{1}{(\rho + p)}\left(\frac{\zeta}{2\tau_\zeta} +
\frac{\kappa T}{\tau_\kappa} + \frac{2\eta}{3\tau_\eta}\right),\]
where $\zeta$, $\kappa$ and $\eta$ are the transport coefficients
of bulk viscosity, heat conduction and shear viscosity,
respectively and $\tau_i^s$ are the corresponding relaxation
times, is a measure of the strength of expansion during the
inflationary phase. Larger values of $\alpha$ lead to stronger
expansion. Furthermore, more efficient models of inflation can be
constructed by including bulk viscosity, heat conduction and shear
viscosity thus strengthening the case for inhomogeneous
cosmological models.

\section{Causal Thermodynamics}

In order to study the influence of relaxational effects when the
cosmic fluid departs from hydrostatic equilibrium we employ a
causal heat transport equation of Maxwell-Cattaneo form given by\cite{yroy, is1, is2, jouu}
\begin{equation} \label{causalgen} \tau h_a{}^b \dot{Q}_b+Q_a =
-\kappa \left( h_a{}^b\nabla_b T+T\dot{u}_a\right)
\end{equation}
where $h_{ab}=g_{ab}+u_a u_b$ projects into the comoving rest
space, $T$ is the local equilibrium temperature, $\kappa$
($\geq0$) is the thermal conductivity, and $\tau$ ($\geq 0$) is
the relaxational time-scale which gives rise to the causal and
stable behaviour of the theory. The noncausal Fourier heat
transport equation is obtained by setting $\tau = 0$ in
(\ref{causalgen}). For the metric (\ref{a1}), equation
(\ref{causalgen}) becomes
\begin{equation} \label{causal}
\tau{(QA)}\!\raisebox{2mm}{$\cdot$}  + QA = -\frac{\kappa
(T)'}{A}\,
\end{equation}where $T'$ represents the temperature gradient.
Note that on setting $\tau = 0$, we regain the Eckart heat
transport equation
\begin{equation}
QA = -\frac{\kappa (T)'}{A}\,
\end{equation}
which was utilised by Singh and Beesham to obtain noncausal
temperature profiles. Following Triginer and Pavon\cite{pavvo}, we assume the
thermal conductivity is that for a radiation fluid interacting
with matter
\begin{equation}
\kappa = cT^3 \sigma,\end{equation} where $c > 0$ is a constant and
$\sigma$ is the mean collision time. The mean collision time is
related to the particle number density $n$ via
\begin{equation} \label{col1}
\sigma = \alpha n^{-1/3},\end{equation} where $\alpha > 0$ is an
arbitrary constant. The particle conservation equation
\begin{equation} \label{col2}
\frac{dn}{ds} + n\Theta = 0,\end{equation} which yields $n \propto
(AB^2)^{-1}$. We can finally write
\begin{equation}
\sigma = \alpha (AB^2)^{1/3},\end{equation} where $\alpha > 0$ is
another constant. Since $\Theta^{-1}$ is the only natural time
scale of the cosmological fluid, we define the relaxation time as
follows
\begin{equation} \label{col3}
\tau = \beta|\Theta^{-1}|=\beta\left(\frac{{\dot A}}{A} +
2\frac{{\dot B}}{B}\right)^{-1}\end{equation} where $\beta (\geq 0)$ can be viewed as a causality 'switch'. By setting $\beta = 0$ we regain the noncausal Eckart transport equation. The mean collision time is related to the relaxation time via (\ref{col1}), (\ref{col2}) and (\ref{col3}). We can write
\begin{equation} \label{col4}
\sigma = \sigma_0 e^{-\int{\frac{\beta}{\tau}}dt}\end{equation} where $\sigma_0 >0$ is a constant. Utilising (\ref{23}) and (\ref{24}) in (\ref{col3}) we obtain
\begin{equation}
\tau = \frac{\beta}{3l}
\end{equation}
which corresponds to constant relaxation time. From (\ref{col4}), we can immediately write
\begin{equation}
\sigma = \sigma_0 e^{-3lt}\end{equation}. We can conclude that the assumption made in (\ref{col3}) holds to good approximation in the early evolution of the cosmological fluid when temperatures are sufficiently high \cite{pavvo4}.
The causal heat
transport equation (\ref{causal}) becomes
\begin{equation}
\beta\left(\frac{{\dot A}}{A} + 2\frac{{\dot
B}}{B}\right)^{-1}{(QA)}\!\raisebox{2mm}{$\cdot$} + Q{A} =
-c_0T^3\frac{(AB^2)^{1/3}}{A}T',\end{equation} which easily
integrates to

\begin{equation}\label{temperr}
T^4
=-\frac{4}{c_0}\left(\frac{A}{B}\right)^{2/3}\left[\beta\left(\frac{{\dot
A}}{A} + 2\frac{{\dot
B}}{B}\right)^{-1}{(QA)}\!\raisebox{2mm}{$\cdot$} + Q{A}\right]x
 +
{\cal {F}}(t),\end{equation} where
\begin{equation}
Q = \frac{2}{A^2}\left[\frac{\dot B}{B} - \frac{\dot
A}{A}\right]\end{equation} The above equation is readily solved to
give us the noncausal temperature profiles for the two cases that
we have investigated thus far. For $n=0$
which corresponds to the case of constant deceleration parameter $
q = -1$ yields
\begin{equation} \label{temptemp}
T^4 = \frac{-24k_1}{c_0c^4}\exp(-4lt)\left[\beta
\left(\frac{k_1}{c^3}\exp(-3lt)-2\right)+1\right]x+{\cal {F}}_2(t)
\end{equation}
where ${\cal F}_2(t)$ is a function of integration. Note that (\ref{temptemp}) does not guarantee $T > 0$. Requiring $T > 0$ on physical grounds will constrain the free parameters appearing in (\ref{temptemp}). The noncausal temperature is obtained by setting $\beta = 0$ in (\ref{temptemp}). This would imply (from (\ref{col3})) that thermal equilibrium is achieved instantaneously which is one of the pathologies of the Eckart theory. It has been
pointed out that entropy of the Universe behaves like a an
ordinary system and tends to a maximum value of the order of
$H^{-2}$ as $a \rightarrow \infty$ \cite{pavvo1}. The rate of
entropy production is given by
\[
S^a_{;a} = \frac{Q^aQ_a}{T^2} =
\frac{4(d_1)^{4/3}k_1^2\exp[-\frac{4k_1\exp[-3lt]}{3c^3l} -
8lt]}{c^8\sqrt{-24\frac{c^4k_1x\exp[-4lt](1 + (-2 +
k_1\exp[-3lt]/c^3)\beta)}{c_0}} + {\cal F}_2(t)},\] which vanishes
as $t \rightarrow \infty$ for an appropriate choice of ${\cal F}$.
Let us consider the temperature profile for the case $n = 0$.
Figures 1 and 2 show the evolution of the causal and noncausal
temperature profiles respectively. In order to generate these
plots we chose the following parameters: $d_1 = 1, k_1 = 10000,
c=0.001, l = 1, c_0 = -1$. Figure 1 corresponds to the case $\beta
= 0$ giving the noncausal temperature. Figure 2 corresponds to the
case $\beta = 1000$ representing the causal temperature profile.
It is evident that the causal temperature is everywhere greater
than its noncausal counterpart. In the infinite past both the
causal and noncausal temperatures are at a maximum and decrease as
the fluid evolves with time. The drop-off in the temperature is
greater in the causal case than the noncausal case indicating that
cooling is enhanced by relaxational effects.

\section{Concluding remarks}

We have successfully obtained the causal temperature profile
for an LRS Bianchi type-V cosmological fluid with constant
deceleration parameter. Our results generalise the thermodynamical
results obtained by Singh and Beesham \cite{bee1}. Our
investigation show that relaxational effects within the cosmic
fluid leads to a higher temperature . We also found that the rate
of entropy production decreases as the Universe evolves in time,
tending to zero for late times. It would be interesting to
investigate the evolution of the temperature profile for the LRS
Bianchi type-V universe with dissipation by employing a full
causal heat transport equation. Work in this direction has been
initiated.

\section*{Acknowledgments}

The authors are thankful to the comments and suggestions made by the anonymous referee which helped clarify some of the results of this paper.

\newpage
\begin{figure}
\centering
\includegraphics[scale=0.6]{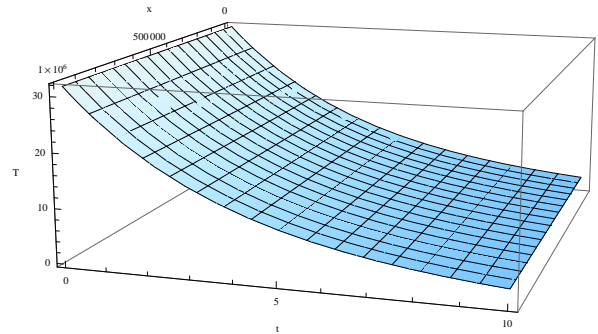}\caption{Noncausal temperature as a function of the radial $x$ and temporal $t$ coordinates} \label{fig1}
\end{figure}

\begin{figure}
\centering
\includegraphics[scale=0.6]{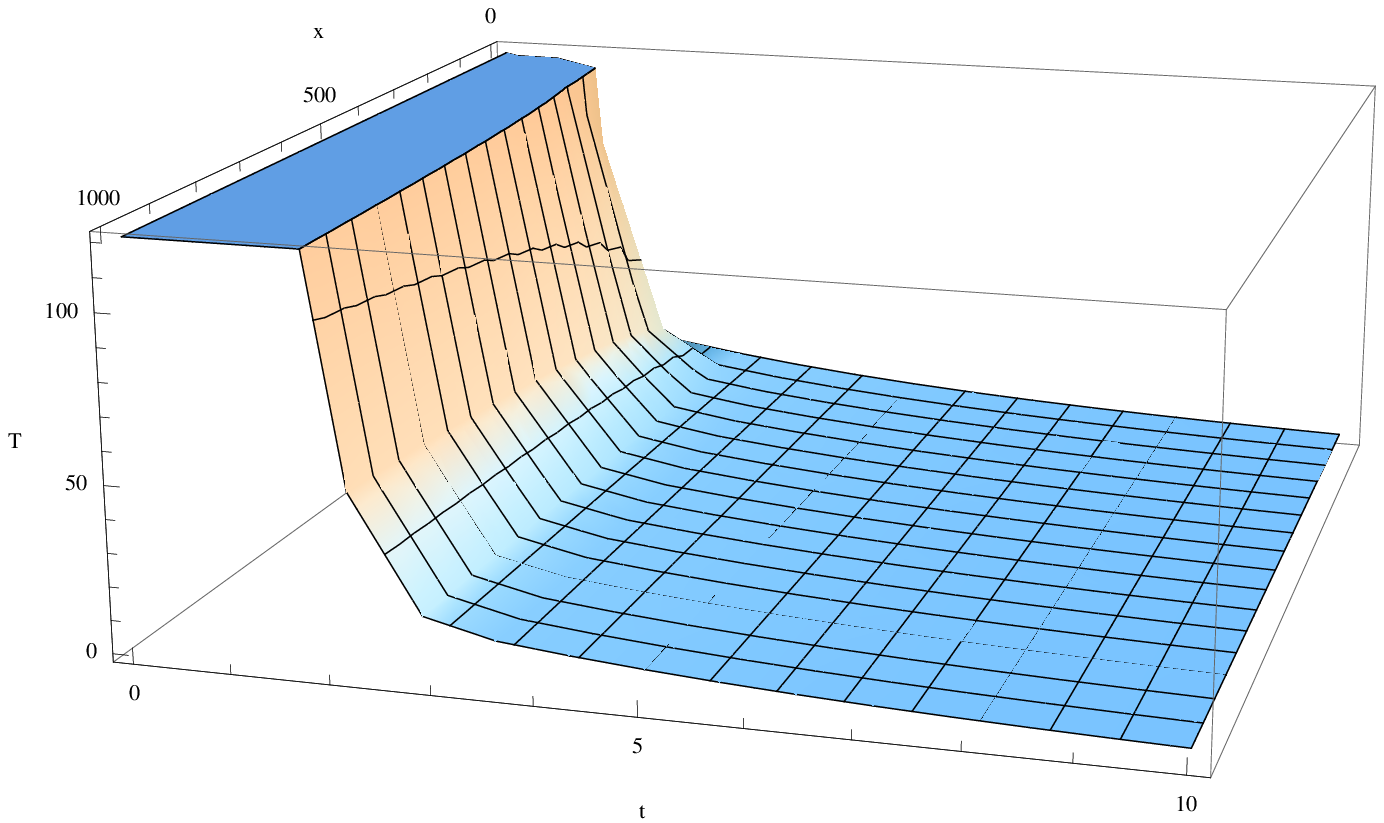}\caption{Causal temperature as a function of the radial $x$ and temporal $t$ coordinates} \label{fig2}
\end{figure}
\end{document}